# Magnetic and structural characterization of nanosized BaCo$_x$Zn$_{2-x}$Fe$_{16}$O$_{27}$ hexaferrite in the vicinity of spin reorientation transition

**A Pasko[1], F Mazaleyrat[1], M LoBue[1], V Loyau[1], V Basso[2], M Küpferling[2], C P Sasso[2] and L Bessais[3]**

[1] SATIE, ENS Cachan, CNRS, UniverSud, 61 av President Wilson, F-94235 Cachan, France
[2] INRIM, Strada delle Cacce 91, I-10135 Torino, Italy
[3] ICMPE, CNRS, Université Paris 12, 2-8 rue Henri Dunant, F-94320 Thiais, France

E-mail: pasko@satie.ens-cachan.fr

**Abstract**. Numerous applications of hexagonal ferrites are related to their easy axis or easy plane magnetocrystalline anisotropy configurations. Certain W-type ferrites undergo spin reorientation transitions (SRT) between different anisotropy states on magnetic field or temperature variation. The transition point can be tuned by modifying the chemical composition, which suggests a potential application of hexaferrites in room temperature magnetic refrigeration. Here we present the results of structural and magnetic characterization of BaCo$_x$Zn$_{2-x}$Fe$_{16}$O$_{27}$ ($0.7 \leq x \leq 2$) doped barium ferrites. Fine powders were prepared using a sol-gel citrate precursor method. Crystal structures and particle size distributions were examined by X-ray diffraction and transmission electron microscopy. The optimal synthesis temperature ensuring complete formation of single W-phase with limited grain growth has been determined. Spin reorientation transitions were revealed by thermomagnetic analysis and AC susceptibility measurements.

## 1. Introduction

Hexaferrites have been the subject of intensive studies due to an appealing combination of good magnetic properties and low cost. This large family of oxides with hexagonal crystal structure contains ferrimagnetic compounds with easy axis of magnetization (*e.g.* M-type ferrites) and easy plane of magnetization (*e.g.* Y-type ferrites). Hence, hexaferrites have been widely adopted in two distinct fields: permanent magnets and microwave technology components [1]. On the other hand, W-type ferrites Ba$M_2$Fe$_{16}$O$_{27}$ ($M$ = Mg, Mn, Fe, Co, Ni, Cu, Zn) can undergo spin reorientation transitions (SRT) between different anisotropy configurations (easy plane ↔ easy cone ↔ easy axis) induced by change of temperature or applied magnetic field [2–9]. The transition temperatures can be tuned by modifying the chemical composition (substitution of bivalent metal $M$). Moreover, some SRT are expected to be of the first order, which suggests a potential application of W-type ferrites in room temperature magnetic refrigeration [10].

    Conventional ceramic method of hexaferrite synthesis is efficient, but requires elevated temperatures for solid state reaction to occur between premixed powders. Alternative production routes (aerosol pyrolysis, chemical co-precipitation, glass crystallization, hydrothermal synthesis, *etc.*)

are intended to improve mixing of initial components down to atomic level and thereby to facilitate diffusion. In particular, a sol-gel technique enables to obtain sufficiently homogeneous precursors for low-temperature synthesis of nanosized simple (spinel) ferrites. However, hexagonal ferrites with complex layered crystal structures still require relatively high temperatures to form because of thermodynamic stability conditions. Trying to avoid rapid growth of grains, we have used this soft-chemistry approach for production of W-type ferrites. An important goal was to determine the heat treatment regimes ensuring complete transformation of a precursor into the smallest particles of single W-phase. In this paper the results of structural and magnetic characterization of $BaCo_xZn_{2-x}Fe_{16}O_{27}$ powders are presented. The chemical compositions were chosen so that SRT occur near room temperature ($x = 0.7, 0.75, 0.8$) or much higher ($x = 2$).

## 2. Experimental

Hexaferrite powders have been prepared by a sol-gel citrate precursor method [11–13]. High purity iron(III) nitrate, barium hydroxide, cobalt(II) oxide, zinc oxide and citric acid were used as starting materials with the molar ratio of citrate to metal ions 2:1. Iron(III) nitrate was dissolved in deionized water and quantitatively precipitated with excess of ammonia solution as iron(III) hydroxide. The precipitate was filtered and washed with water until neutrality. Then the obtained iron(III) hydroxide was dissolved in a vigorously stirred citric acid solution at 60–70 °C. Barium hydroxide and other metal sources were added according to stoichiometry. At this stage 3 samples of each composition were separated: (A) pH value of the solution was adjusted to 6 for better chelation of cations [13]; (B) no modification was done [12]; (C) ethylene glycol was added to increase viscosity by polycondensation reaction [11]. Water was slowly evaporated at 80–90 °C with continuing stirring until a highly viscous residue is formed. The gel was dried at 150–170 °C and heat treated for 2 h at 450 °C for total elimination of organic matter. Finally, the inorganic precursor with homogeneous cationic distribution was calcined at temperatures up to 1300 °C for 2 h with heating/cooling rate 200 K/h to synthesize a hexaferrite phase. Details of the procedure will be discussed elsewhere.

Crystal structures were examined by PANalytical X'Pert Pro X-ray diffractometer (XRD) in Co-K$_\alpha$ radiation with X'Celerator detector for rapid data acquisition. Magnetization curves were recorded on Lake Shore vibrating sample magnetometer (VSM). Direct observations of powder particles were carried out by FEI Tecnai G$^2$ F20 transmission electron microscope (TEM) operating at 200 keV. Thermogravimetric analyzer (TGA) PerkinElmer Pyris 6 equipped with a permanent magnet was used for thermomagnetic measurements above room temperature. AC magnetic susceptibility at low temperatures was studied in Quantum Design PPMS. Rietveld analysis of XRD spectra was performed using MAUD software [14].

## 3. Results and discussion

Characterization had a twofold purpose: to find a correlation between the calcination temperature and the final product properties (phase composition, mean grain size, coercivity); to confirm the existence of SRT at expected temperatures in fine hexaferrite powders.

*3.1. Structural characterization and phase analysis*

Figure 1 represents XRD patterns from $BaCo_2Fe_{16}O_{27}$ powder synthesized at different temperatures. W-ferrite becomes the major phase at 1300 °C, while lower calcination temperatures lead to a mixture of W-ferrite, M-ferrite (both are hexagonal) and S-ferrite (with spinel structure) in the final product. Traces of $\alpha$-$Fe_2O_3$ are sensitive to the method of preparation and dissolve completely with increase of calcination temperature; however, other secondary phases may appear together with W-ferrite. The results of quantitative phase analysis based on Rietveld method of XRD full profile fitting are shown in figure 2. At 900 °C only M-ferrite can be formed, with excess of cobalt giving also S-ferrite. Increase of temperature creates favorable conditions for W-ferrite synthesis, directly from the precursor components (nanocrystalline or amorphous oxides) or through the intermediate reaction

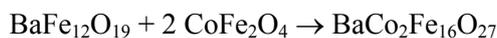

$$BaFe_{12}O_{19} + 2\ CoFe_2O_4 \rightarrow BaCo_2Fe_{16}O_{27}$$

The proportion between 3 ferrite phases depends on calcination temperature and production route (as the precursor composition is affected, for instance, by pH value of the solution). However, at 1300 °C M-ferrite and S-ferrite almost completely disappear.

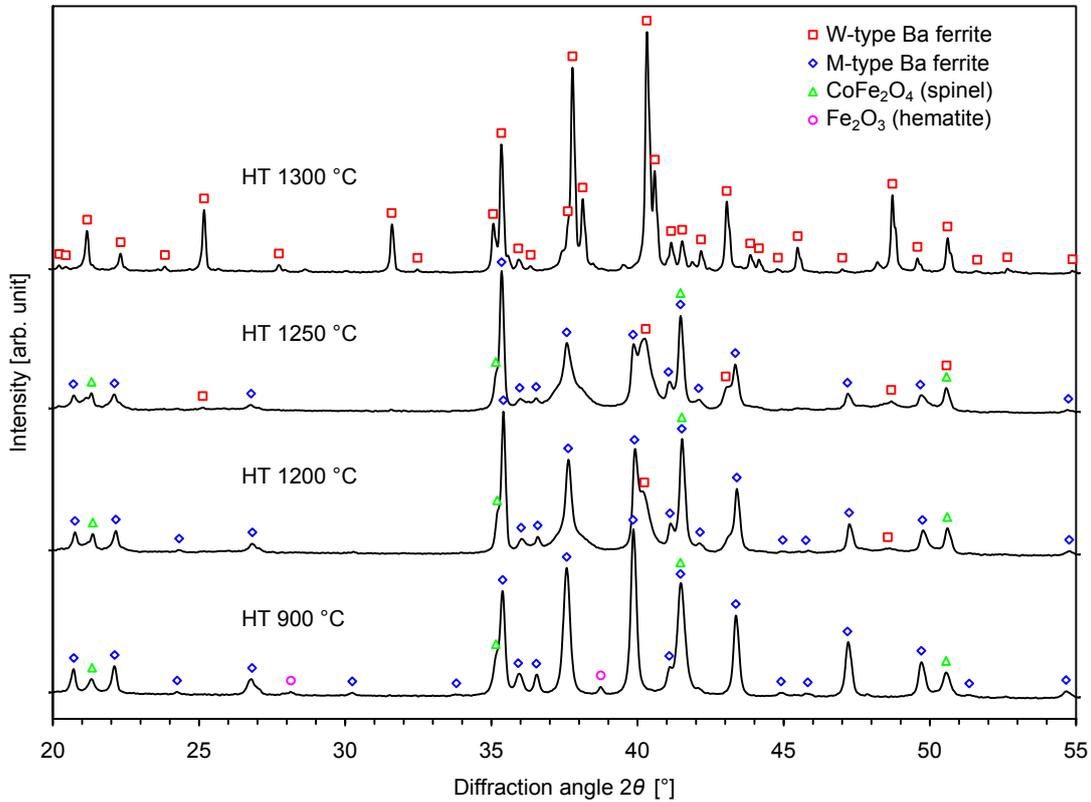

**Figure 1.** XRD patterns of $BaCo_2Fe_{16}O_{27}$ powder A taken after different heat treatments.

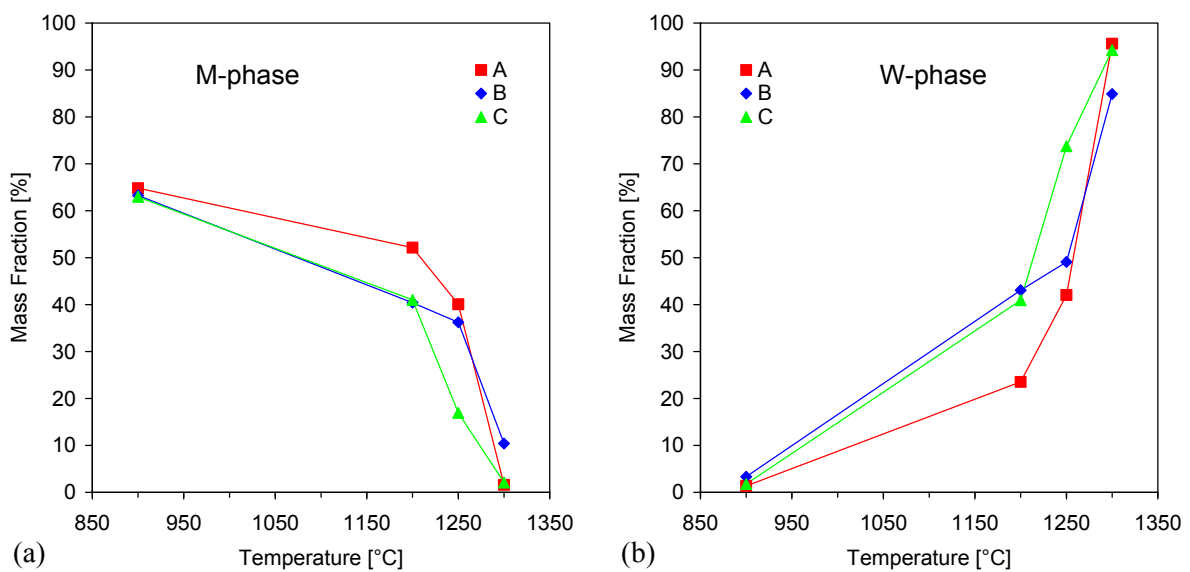

**Figure 2.** Phase composition of $BaCo_2Fe_{16}O_{27}$ powders A, B, C (from Rietveld fitting of XRD spectra) *vs.* calcination temperature: (a) M-ferrite fraction; (b) W-ferrite fraction.

## 3.2. Morphology and average size of particles

The size of diffracting crystallites in W-ferrite powders is derived from XRD data by analysis of line broadening using Rietveld refinement as shown in figure 3(a). It gives a minimum estimate for average particle size (a particle can contain several crystallites). The mixture of M-ferrite and S-ferrite formed at 900 °C have a mean size of crystallites ~ 50 nm. With increase of calcination temperature the calculated values slightly decrease, this can reflect contribution to line broadening caused by generation of defects in the transforming mixture of phases. At 1300 °C, when W-ferrite phase rapidly grows and becomes dominant, the apparent size of crystallites significantly increases. In addition, XRD spectra of the powders synthesized at this temperature exhibit a specific texture characteristic for plate-like particles pressed in a holder.

Magnetic measurements give an illustration to these phase transformations. Figure 3(b) shows that after calcination at 900 °C coercivity is high due to the presence of M-ferrite. When W-ferrite forms, the coercivity decreases and reaches ~ 15 mT at 1300 °C, a typical for this phase value.

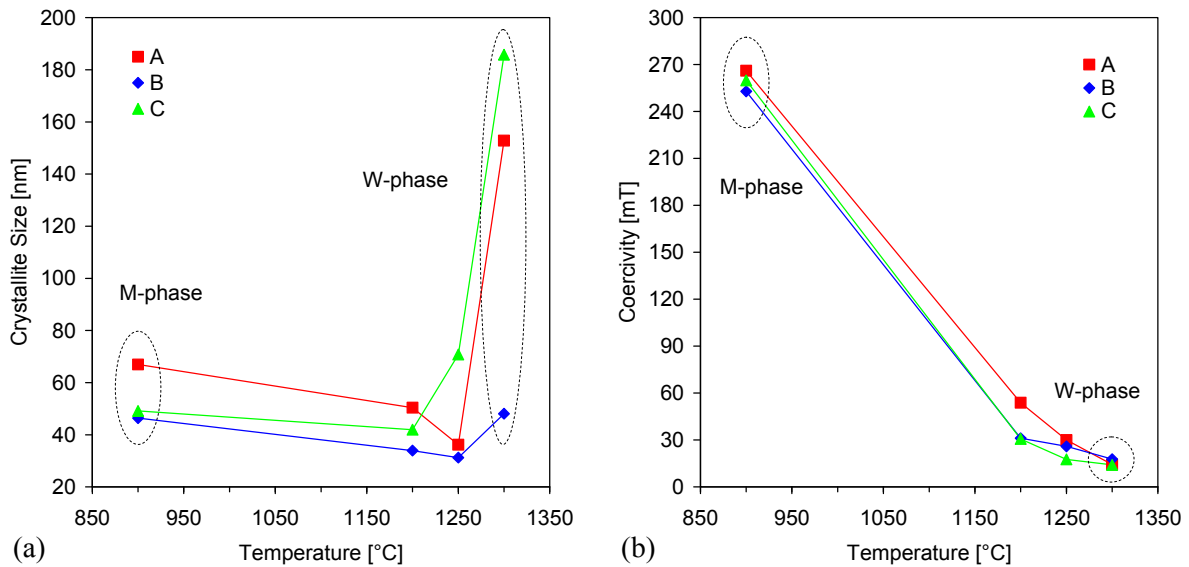

**Figure 3.** $BaCo_2Fe_{16}O_{27}$ powders A, B, C calcined at different temperatures: (a) mean size of crystallites (from XRD line broadening); (b) coercivity (from VSM measurements).

Direct observations of W-ferrite particles were performed by TEM. Powders calcined at 900 °C consist of aggregates of clean and almost round particles not exceeding ~ 100 nm, an example is shown in figure 4(a). The shape of crystallites reflects their hexagonal crystal structure. Powders calcined at temperatures high enough to form W-ferrite exhibit absolutely different morphology as shown in figure 4(b). The particles have a wide size distribution from 50 nm to 2000 nm and are mostly plate-like. This morphology is observed not only for 1300 °C when W-ferrite is almost single phase, but also for lower temperatures when W-ferrite appears in the precursor. The average particle size of $BaCo_xZn_{2-x}Fe_{16}O_{27}$ ($x = 0.75$) powder estimated by laser light scattering analysis is ~ 700 nm, which agrees with TEM observations.

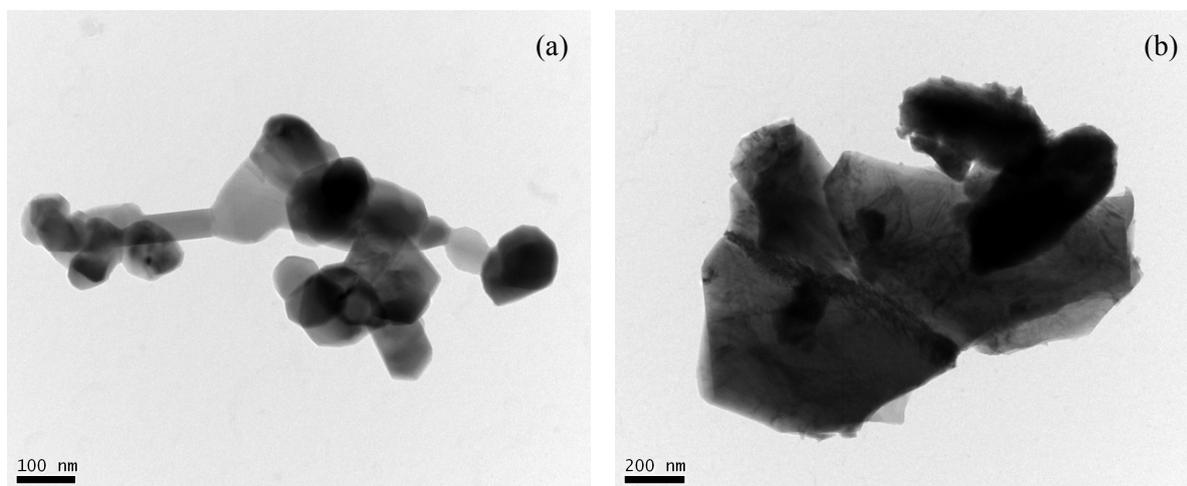

**Figure 4.** Typical TEM images of powder particles: (a) $BaCo_2Fe_{16}O_{27}$ precursor after calcination at 900 °C (mixture of ferrites); (b) W-type $BaCo_xZn_{2-x}Fe_{16}O_{27}$ ($x = 0.75$) ferrite synthesized at 1275 °C.

### 3.3. Spin reorientation transitions

Thermomagnetic measurements were performed on loose powders in a weak (1–10 mT) magnetic field with cycling from room temperature up to 600 °C. An example of magnetization as a function of temperature for W-type $BaCo_xZn_{2-x}Fe_{16}O_{27}$ ferrite with $x = 0.8$ is shown in figure 5(a). Heating and cooling segments do not match (typical hysteretic behaviour), and the first heating differs from others because of particular magnetization history. Magnetic susceptibility is higher in the easy plane state than in the easy axis one, therefore both spin reorientation (at ~ 55 °C) and ferrimagnetic-paramagnetic (at ~ 430 °C) transitions are well resolved. Small amplitude of the anomalies observed above W-ferrite Curie point confirms that fractions of magnetic secondary phases are negligible.

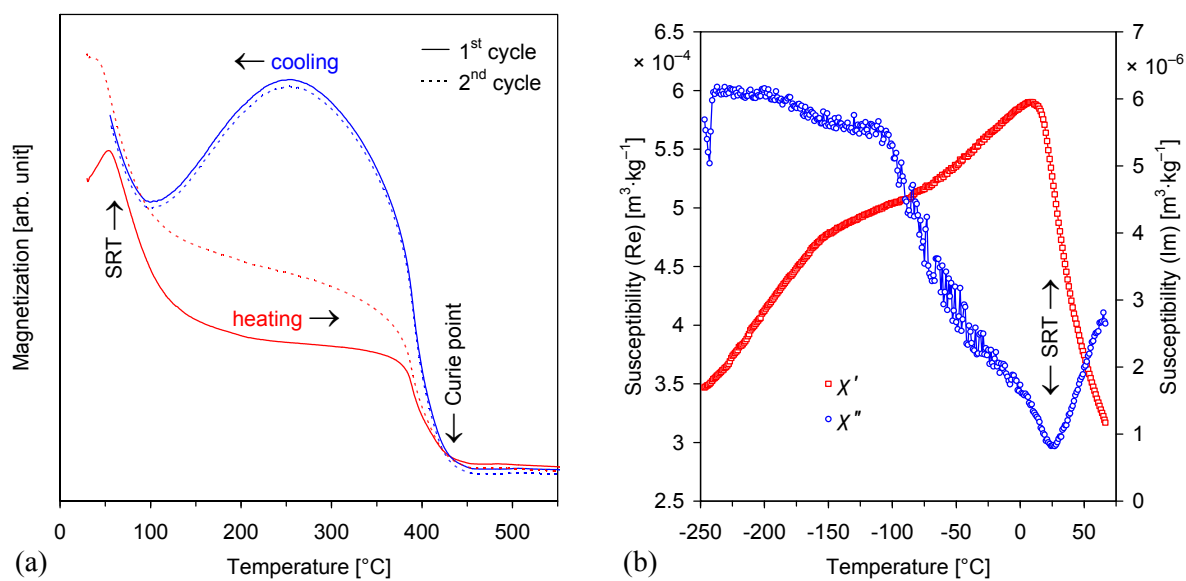

**Figure 5.** Magnetic transitions in W-type $BaCo_xZn_{2-x}Fe_{16}O_{27}$ ferrites: (a) magnetization measurements for $x = 0.8$ (higher temperatures); (b) AC susceptibility measurements for $x = 0.7$ (lower temperatures).

On the first heating, a sharp drop of magnetization related to SRT is preceded by a distinct peak. The spin reorientation temperature of polycrystalline powder is also characterized by inflection point of magnetization dependence [15]. Another reference can be the point where subsequent cooling and heating curves converge (clearly seen in figure 5(a) on the left of magnetization hump). Unlike [6], the latent heat of SRT has not been reliably detected by differential scanning calorimetry (DSC), while Curie point is well visible.

AC magnetic susceptibility as a function of temperature for W-type BaCo$_x$Zn$_{2-x}$Fe$_{16}$O$_{27}$ ferrite with $x = 0.7$ is shown in Fig. 5(b). The real part of susceptibility has a maximum at ~ 6 °C followed by a sharp drop attributed to SRT. Its imaginary part has a minimum at ~ 27 °C which can also be taken as the spin reorientation temperature. Moreover, the minimum of imaginary part turns out to approach the inflection point of the real part.

## 4. Conclusions

Single-phase W-type ferrite submicron powders with different magnetocrystalline anisotropy configurations are synthesized through a sol-gel citrate precursor route. The effect of calcination temperature and other production parameters on the phase composition, average particle size and magnetic properties of the powders is established. Chemical compositions of substituted hexaferrites undergoing spin reorientation transitions near room temperature are determined.

**Acknowledgements**
This work is supported by EC Seventh Framework Programme funding (project SSEEC, contract FP7-NMP-214864). The authors are grateful to P. Audebert (ENS Cachan) for assistance in sol-gel technique and G. Wang (ICMPE) for TEM observations.